# Role of phytochemicals in the chemoprevention of tumors

Catalano E[1]

1- Università degli Studi di Bari Aldo Moro, Via Edoardo Orabona, 4, 70126 Bari

**Abstract**

Phytochemicals are plant-derived secondary metabolites, which may exert many biological activities in humans, including anticancer properties. Although recent findings appear to support their role in cancer prevention and treatment, this issue is still controversial. Anti-cancer activity of phytochemicals mainly depends on their multi-target mechanism of action, including antimutagenic, antioxidant and antiproliferative activities. Furthermore, they may modulate the host immune response to cancer, reducing inflammatory microenvironment and enhancing lymphocyte onco-surveillance. Since carcinogenesis is multi-factorial and involves several signaling pathways, multi-targeted agents as phytochemicals may represent promising anticancer compounds. This narrative review aims to analyze the current literature on phytochemicals highlighting their specific targets on carcinogenic molecular pathways and their chemopreventive role. A full comprehension of their activity at molecular and cellular levels will contribute for a better understanding of phytochemical clinical efficacy, thus promoting the identification of new effective plant-derived therapeutics.

## 1. Introduction

Overwhelming evidence from epidemiological, in vivo, in vitro, and clinical trial data suggests that the plant-based diet can reduce the risk of chronic diseases (e.g., cardiovascular disease, hypertension, diabetes, and cancer) due to presence of biologically active plant compounds or phytochemicals. Phytochemicals include those plant-derived compounds that have specific biological activity in human [1]. They can be defined also as bioactive natural molecules that can be of benefits for human health [2].

In fact single phytochemicals and enriched natural extracts able to interfere with self-renewal and drug resistance pathways in cancer cells were investigated. This is a milestone in the improvement of cancer treatment because the synthetic anticancer drugs that are currently used are often highly toxic for healthy organs and weakens the patient's immune system.

Indeed, in terms of prevention, their beneficial effects have been described in different epidemiological investigations, which underlined, despite the limitations of these kind of studies, a reliable relationship between diets rich in phytochemical and reduction in the risk of developing several diseases, including cancer [3]. Indeed, in particular, high intake of fruits and vegetables, the richest dietary components in phytochemicals, has been correlated to a decrease in the risk of several cancers [4]. Nevertheless, one-third of all cancer deaths is estimated to be preventable by "healthy" lifestyles, including appropriate nutrition [4]. A plethora of phytochemicals, such as carotenoids, antioxidative vitamins, phenolic compounds, terpenoids, steroids, indoles, and fibers, has been considered responsible for the risk reduction [5].

On the other hand, phytochemicals can be assimilated to chemotherapeutics, which are frequently derived from natural substances, directly extracted from plants, or other natural sources, or chemically derived from naturally occurring compounds [2]. Moreover, behind traditional medicine, a large number of cancer patients are currently using plant-derived compounds in the context of complementary therapies [2].

Therefore, a growing interest is arising around phytochemicals role in cancer prevention and treatment. Recent research suggests that the investigation of "new" phytochemicals and related molecular targets can be exploited to identify novel anti-cancer drugs, following sequential steps. This approach consists in the preliminary selection of phytochemical candidates for cancer prevention or therapy, basing on the pre-clinical results related to cell-transformation and anti-tumorigenic activity assays. Phytochemicals need to be further validated by means of *in vivo* models, determining pharmacokinetics and pharmacodynamic molecular interactions and targets. Clinical trials should assess anti-cancer efficacy, further investigating specific pharmacokinetics and pharmacodynamics in humans. Phytochemicals can contribute to cancer prevention by influencing different stages of the tumor development, from tumor initiation through all the phases of cancer [6], such as cell proliferation, apoptosis, invasion and metastasis, angiogenesis and immortality [7].

From this perspective, this review has the aim to summarize the current body of evidence dealing this particular issue, focusing on specific phytochemicals possessing anticancer properties and their molecular targets, also suggesting new strategies for further biomedical application and future directions.

**Phytochemicals anti-cancer properties**

Natural phytochemicals were classified according to their chemical structure, botanical origin, biological properties and biosynthesis. In this review, the phytochemicals with antitumor activities will be described, concerning the pathways and cancer subtypes which they target (Table 3) [10]. In the last years chemoprevention by phytochemicals has focused a great attention and is considered to be a feasible, readily applicable, acceptable, and accessible approach to cancer control, regression and management. In fact several phytochemicals are in preclinical or clinical trials for cancer chemoprevention. They can disrupt the carcinogenetic process at different level: initiation, promotion progression. In fact, cancer results from this multistage carcinogenesis process that involves 3 distinguishable but closely connected stages: initiation (normal cell → transformed or initiated cell), promotion (initiated cell → preneoplastic cell), and progression (preneoplastic cell → neoplastic cell).

Initiation. Initiation begins with a mutation that reduces the cell's sensitivity to tissue growth constraints regarding the activation of oncogenes and the inhibition of tumor suppressors.

Promotion. Experimental evidence demonstrates that, after initiation, tumor formation is stimulated by an external event such as wounding or inflammation [12]. Different studies demonstrate that the initiation event will not be tumorigenic without introduction of a second environmental perturbation that increases the maximum substrate delivery rate. It allows cellular evolution presumably through the clastogenic effects of hypoxia and acidosis that accompanies wounding and inflammation or by creating a harsh environment that increases selection pressures.

Progression. Although tumor growth continues following initiation and promotion, the basement membrane remains intact (breeching the integrity of the basement membrane defines the transition between premalignant lesions and invasive cancer). Since blood vessels remain deep to the basement membrane, substrate must diffuse over increasingly long distances resulting in severe hypoxia in those cells >100 microns (about 5 cell layers) from the membrane (13,14). The following steps are involved in cancer progression: accelerated passage of cells through checkpoints within the cell cycle, impaired responses to normal apoptotic signals or to other stimulators of programmed cell death, overproduction of growth regulatory hormones, enhanced metastasis of cancerous lesions, and alteration in host immune responses. These altered functions have a central underlying biochemistry in that they all involve cellular signal transduction.

Noteworthy, phytochemicals have been recently introduced in cancer therapy as "adjuvant therapy". In this perspective adjuvant therapy is defined like an additional cancer treatment given after the primary treatment to lower the risk that the cancer will come back. The aim of adjuvant therapy is also to slow cancer development. Generally adjuvant therapy may include chemotherapy, radiation therapy, hormone therapy, targeted therapy, or biological therapy. In addition phytochemicals may play a major role in supporting chemotherapy drugs, and primarily target proliferating tumor cells.

**Therapeutic applications of phytochemicals**

Phytochemicals may be tested as potential chemopreventive or chemotherapeutic agents. The difference existing between primary and secondary chemoprevention and chemotherapy has been previously defined [2,144,145]. Chemoprevention corresponds to a therapeutic preventive treatment to be administered to "healthy" people who agree to receive phytochemicals or natural bioactive compounds to prevent the cancer onset. A key issue in chemoprevention is lack of toxicity or undesirable side effects of substances administered. Phytochemicals to be applied in chemoprevention should respect the properties as follows: determination of the adequate dose of phytochemicals which is not neutralized by metabolizing enzymes, and, in parallel, is not too high to cause toxicity in humans. Moreover, oral administration of the active principle as pills, capsules or similar formulations should be better and more appropriate for chemoprevention respect to other modalities. On the opposite chemotherapy corresponds to treatments for patients who already developed tumors and/or invasive cancers. In this case, drugs are normally given at pharmacological doses and their intrinsic toxicity is related to a strict benefit-versus-risk analysis. Key issues of phytochemical chemotherapy are as follows: modality of phytochemicals administration (e.g., by mouth or injected into the blood), number of phytochemical doses administered and their dosage of safety. Phytochemicals, which are administered per os, can be found in free form in the blood only if they are taken at pharmacological doses (hundreds of milligrams). A crucial aspect concerns the molecular pathways involved in the metabolism of phytochemicals that corresponds to the process of conjugation (methylation, sulfatation and glucuronidation). Phytochemicals to be used for chemotherapy should be potentially bioavailable and biologically active [146]. An alternative strategy to increase the bioavailability and biodistribution of phytochemicals is their administration by intravenous injection to avoid the formation of conjugates. The correct dose of phytochemicals is crucial to determine the primary site of metabolism, with high doses primarily metabolized in the liver and low doses in the intestine. In the last years only few phytochemicals were employed for clinical trials as chemotherapeutic agent against specific types of cancers such as e.g., curcumin which reached phase III clinical trial. Another approach to introduce the anticancer efficacy of phytochemicals into clinics is based on the possibility of combination therapy, where anticancer compounds are given in association with well-known drugs currently used in chemotherapy. From a pharmacologic point of view, this strategy presents several advantages. Phytochemicals are functionally pleiotropic: we clearly showed that they possess multiple intracellular targets, affecting different cell signaling processes usually altered in cancer cells, with limited toxicity on normal cells. Targeting simultaneously multiple pathways may help to kill cancer cells and slow drug resistance onset. In addition, if proven, the association of pure or synthetic analogs of phytochemicals with chemotherapy or radiotherapy may take advantage of the synergic effects of the combined protocols, resulting in the possibility to lower doses and, consequently, reduce toxicity. The results of the ongoing clinical trials could provide good results for designing future large-scale clinical trials to ascertain the full chemopreventive and chemotherapeutic efficacy of phytochemicals.

## 2. Anti-cancer phytochemicals

### 2.1 Targeting molecular pathways

Phytochemicals are functionally pleiotropic: they trigger pathways linked to different pathophysiological conditions and possess multiple intracellular targets, affecting different cell signaling processes usually altered in cancer cells, with limited toxicity on normal cells. Targeting simultaneously multiple pathways may help to kill cancer cells and slow drug resistance onset. In addition, if proven, the association of pure or synthetic analogs of phytochemicals with chemotherapy or radiotherapy may take advantage of the synergic effects of the combined protocols, resulting in the possibility to lower doses and, consequently, reduce toxicity. Phytochemicals express their anti-cancer activity mainly by blocking mutagenic activity of other compounds, reducing damage induced by oxidative stress (DNA, protein oxidation and lipid peroxidation) and suppressing abnormal cell proliferation [8]. Further properties include modulation of immune response, in terms of reduced inflammation and lymphocyte onco-surveillance [9].

The molecular targets and anti-cancer activities of phytochemicals are summarize in Table 2, supporting as phytochemicals especially inhibit carcinogenesis-related genetic and metabolic pathways [10]. In fact, phytochemicals have multiple molecular targets with antitumor activity that are capable of interfering in every stage of cancer development or process: cell proliferation, replicative immortality, oxidative stress, cell death evasion, angiogenesis, invasion & metastasis, immune evasion, inflammation, cell metabolism alteration, genome instability.

**Cell proliferation**

The development of cancer within animals usually involves multiple molecular events or steps. These steps often increase cell proliferation within tissues. In particular phytosterols show numerous antitumor effects on transduction signaling pathways that regulate cell proliferation and apoptosis. Phytosterols are derivatives of the parent molecule 4-desmethyl sterol. The most common ones are β-sitosterol, campesterol, and stigmasterol, and they are mainly found in nuts, whole grains and seeds.

*β-sitosterol* - β-sitosterol has a chemical similarity to cholesterol. The main characteristic of this compound is the *anti-proliferative* role against several cancer types. The effects of β-sitosterol were investigated on colon and breast tumors, leukemia, fibro-sarcoma, stomach and prostate cancers. β-sitosterol inhibits tumor cell proliferation and induces apoptosis, especially by activating caspase 3 and 8, increasing Fas levels and MAPK activity [40], and modulating the expression of molecules related to apoptosis, such as Bcl-2 [41]. The pro-apoptotic signal is associated to the intracellular accumulation of ceramide, whom *de novo* synthesis is promoyed by β-sitosterol [42]. Finally, β-sitosterol was found to sensitize breast cancer cells to TRAIL-induced apoptosis [43].

Stigmasterol is an unsaturated plant sterol occurring in the plant fats or oils of soybean, calabar bean, and rape seed, and in a number of medicinal herbs, including the Chinese herbs Ophiopogon japonicus, in Mirabilis jalapa [2] and American Ginseng.

Campesterol is a phytosterol whose chemical structure is similar to that of cholesterol. Many vegetables, fruits, nuts and seeds contain campesterol, but in low concentrations. Banana, pomegranate, pepper, coffee, grapefruit, cucumber, onion, oat, potato, and lemon grass (citronella) are few examples of common sources containing campesterol at ~1–7 mg/100 g of the edible portion

**Oxidative stress and redox signaling**

Phytochemicals can directly inhibit all phases of carcinogenesis in *in vitro* and *in vivo* models acting on redox signaling.5,10,15,16 Phytochemicals can interact with different antioxidant enzymes18,19 such as glutathione S-transferases (GST) and NADPH: quinine oxidoreductase (NQO1). Moreover, phytochemicals can prevent the initiation phase of carcinogenesis by modulation of cytoprotective enzyme activation, like the transcription factor NF-E2-related factor 2 (Nrf2), that is induced not only in the presence of oxidative stress but also in the presence of various phytochemicals. Activation of Nrf2 target genes by phytochemicals is advantageous to maintaining genomic integrity and reducing the effects of electrophiles, chemical challenges, and oxidative stress on DNA.20 Various phytochemicals such as isothiocyanate and sulphoraphane found in cruciferous vegetables and the phytochemicals found in green tea, turmeric and milk thistle act as potent activators of Nrf2 in both cell culture and animal models [5,20,21]. In addition these phytochemicals inhibit the conversion of procarcinogens to DNA damaging species [22]. Thus, phytochemicals reduce the potential for carcinogenic initiation, by behaving as blocking agents that prevent DNA mutations.

**Genome instability**

Genomic instability refers to an increased tendency of alterations in the genome during the life cycle of cells. It is a major driving force for tumorigenesis. During a cell division, genomic instability is minimized by four major mechanisms: high-fidelity DNA replication in S-phase, precise chromosome segregation in mitosis, error free repair of sporadic DNA damage, and a coordinated cell cycle progression. In this context curcuminoids contribute to these mechanisms of prevention of genomic instability and tumorigenesis.

*Curcuminoids*

Curcuminoids are obtained from turmeric as a yellow crystalline powder [52]. The major curcuminoids present in turmeric are curcumin, demethoxycurcumin, bisdemethoxycurcumin, and the recently discovered cyclocurcumin [52]. These compounds were reported to possess anticancer properties, and curcumin is the most deeply investigated.

*Curcumin* - Curcumin inhibits cell growth in many types of cancers, including some uncommon types, i.e. cholangiocarcinoma, medulloblastoma, and uterine leiomyosarcoma (refs). Curcumin has been shown to be beneficial in all 3 stages of carcinogenesis. In fact, a number of studies showed beneficial effects of curcumin in neoplastic and pre-neoplastic diseases, among the others, multiple myeloma, pancreatic and colon cancers [52, alter refs]. Beneficial effects of curcumin are related to the inhibition of NF-κB and, consequently, induce *anti-inflammatory, antioxidant, and antitumor effects* (refs). Curcumin was shown to be a strong inhibitor of NF-κB activity and its inhibitory effect on NF-κB related pathways often leads to cellular apoptotic response. Nuclear factor kappa B (NF-κB) is a strong mediator of inflammation and, in a majority of systems, supports the pro-proliferative features of cancer cells. The application of various anticancer drugs, cytostatics, triggers signals which lead to an increase in cellular NF-κB activity.

Curcumin acts on multiple molecular targets to inhibit all stages of carcinogenesis (refs) such as: mTOR signaling pathway [53], cell cycle progression (Cyclin D1), proliferation (EGFR), survival pathways (β-catenin), transcription factors such as AP-1 [52], other molecules related to metabolism (HIF-1) and invasion and metastasis (CCL2, MMPs), the modulation of apoptosis-related molecules (caspases and Bcl-2 family) [52], and the upregulation of p53 [54]. Curcumin acts on β-catenin like a molecular target with a chemopreventive action. β-catenin/T-cell factor (TCF)/lymphoid enhancer factor (LEF) signaling is disrupted in many cancer cells, such as those of colorectal cancer, hepatocellular carcinoma, and gastric carcinoma [80-82]. Dysregulated β-catenin/TCF is implicated in cancer progression and poor prognosis. Curcumin has been found to reduce the invasion and subsequent metastasis of cancer cells suppressing the MMP expression and thus blocking the invasiveness of cancer cells [83]. Moreover, curcumin inhibits human papilloma virus oncoproteins, which are involved in the onset of cervical cancer [57]. In addition curcumin is able to slow metastasis progression promoted by matrix metalloproteinase (MMPs), recent studies suggest that curcumin blocks extracellular signaling to and decreased expression of MMP-9 by tumor cells in thyroid, colorectal, pancreatic, ovarian, and numerous other cancer cells, in vitro and in animal models.51,52,55

Interestingly, curcumin has been also related to increased *immune-surveillance*, enhancing T cells effectors against tumor cells [55]. Furthermore, it plays an important role in inhibiting multi-drug resistance influencing the activity of the multidrug-resistance-linked ATP Binding Cassette (ABC) drug transporter, ABCG2 in animal models [56]. ABCG2 protein is involved in drug absorption and/or tissue distribution of drugs. Finally, curcumin reached the phase III of a randomized clinical trial and it was tested for its chemopreventive effect.

**Cancer and inflammation**

Inflammation is a critical component of tumour progression. Many cancers originate from sites of infection, chronic irritation and inflammation. Inflammatory cells play a key-role in the neoplastic process within the tumour microenvironment. Moreover, other signalling molecules of the innate immune system are involved in cancer invasion, migration and metastasis, such as selectins, chemokines and their receptors. These observations could be used for new anti-inflammatory therapeutic approaches to cancer development that could involve phytochemicals like luteolin.

*Luteolin-* Among anti-cancer flavones, luteolin can be found in many medicinal herbs and vegetables such as parsley, celery, pepper, and dandelion. In the past, the plants rich in luteolin were used in Oriental medicine against hypertension, inflammatory disease, and cancer [25]. The whole family of flavones seems to lower the risk of breast cancer, but, in particular, luteolin intake was reported to significantly decrease the incidence of ovarian cancer [26].

Luteolin is available at low cost in large amounts from plants, such as *Reseda luteola* [26]. It can *inhibit cancer cell proliferation*, such as nasopharyngeal and oral squamous cancers, by inhibiting several tumor-related signaling pathways such as Akt/PKB and NF-κB pathways. The Akt/PKB signaling pathway is involved in cell signaling leading to cell survival (blocking apoptosis). NF-κB (nuclear factor kappa-light-chain-enhancer of activated B cells) is a protein complex that controls transcription of DNA. In addition, luteolin can decrease the protein expression of Cyclooxygenase-2 (Cox-2) and Prostaglandin-$E_2$ (PGE2) in macrophage-like cells, thus inducing an anti-inflammatory effect that could potentially be beneficial in cancer therapy [27] because luteolin activates superoxide and hydroxyl radical scavenging. In this way it's limited ROS production that has a pivotal role in the early stages of tumor development. Noteworthy, luteolin is able to *induce apoptosis* in multidrug-resistant cancer cells expressing P-glycoprotein and ATP binding cassette, sub-family G member 2 (ABCG2), without affecting their transport functions [27]. Nonetheless, it appeared to inhibit the expression of CD74 in gastric epithelial cells, which is often expressed in a gastric carcinoma cell line and it is important for adhesion of *Helicobacter pylori* to gastric mucosa [28]. The decreased expression of CD74 can inhibit the infection of gastric cells from this bacterium, thus potentially decreasing the possibility of gastritis and gastric cancer.

A previous study showed the antitumor effect of luteolin correlated with oxidative stress in cancer cells: an intracellular increase in levels of reactive oxygen species (ROS). In fact a proteomic approach based on 2D electrophoresis and Western blot analysis showed the induction of proteins involved in ROS metabolism and subsequent cellular apoptosis, such as peroxiredoxin 6 (PRDX6) and prohibitin (PHB), after luteolin treatment in human hepatoma cells. PRDX6 plays a role in the regulation of phospholipid turnover as well as in protection against oxidative injury, while PHB is a negative regulator of cell proliferation and a tumor suppressor. This finding suggests that luteolin may exert *pro-oxidant activity* in tumor cells [29]. Lastly, luteolin, because of its high lipophilicity, can cross biological membranes, penetrate into human skin and overcome the blood-brain barrier, thus being potentially useful against skin and brain cancers.

**Apoptosis**

Programmed cell death or apoptosis is a highly conserved physiological cell suicide response essential for mammalian homeostasis. Apoptosis plays a key-role in the regulation of cell cycle. Apoptosis involves cascades of enzymatic events including apoptosis-related cysteine proteases called cytosolic aspartate-specific proteases or caspases [52]. Depending upon the triggers of apoptosis and which initiator caspases are involved, apoptotic pathways are termed either extrinsic or intrinsic. In particular lycopene is able to regulate the apoptosis process.

*Lycopene -* Lycopene is the most promising carotenoid with effects related to cancer prevention and therapy. Moreover lycopene plays an important role in regulating hormone action, cell cycle, apoptosis, gap-junction communication and epigenetics, suggesting that their antioxidant activity may not be solely responsible for their anticancer effect [59].

An inverse correlation between lycopene intake and prostate cancer risk was observed [75]. Lycopene administered for 3 months (10 mg/day) in patients with prostate cancer decreased PSA, tumor grade and urinary tract symptoms [60]. Moreover, lycopene is able to inhibit other cancer types (Table 3). Lycopene treatment induces an alteration of serum concentrations of components of the IGF system in both prostate and colorectal cancer patients [61, 62]. In vitro, lycopene is able to modify the functionality of different pathways (Table 3). In epidemiological studies, lycopene reveals to be a promising agent in cancer therapy, especially for prostate cancer. A recent study reported that lycopene administration increased the antitumor activity of docetaxel in castration-resistant prostate tumor models [63]. Furthermore, some studies did not demonstrate an inverse correlation between lycopene consumption and cancer risk [64, 65]. This aspect could be explained, in part, by interindividual variations with regard to genetic background. In fact, polymorphisms in the genes XRCC1 and MNSOD were reported to determine the effect of lycopene regarding prostate cancer risk [66, 67]. Finally, an epidemiological study in elderly Americans indicated that high tomato intake was associated with a 50% reduction in mortality from cancers at all sites, and a case-control study in Italy showed potential protection of high consumption of lycopene from tomatoes against cancers of the digestive tract [75].

**Angiogenesis in cancer**

An important role in the final stages of cancer progression (metastasis) is played by angiogenesis and lymphangiogenesis induced by chemical signals from tumor cells in a phase of rapid growth (Folkman 1971). Generally, the cancer cells without blood circulation can grow up to a maximum of 1-2 mm$^3$ in diameter and then stopped, but grow over 2 mm$^3$ when placed in an area where tumor angiogenesis would be possible. In the absence of vascular support, tumors usually may undergo necrotic or even apoptotic processes (Holmgren et al 1995; Parangi et al 1996). Thus, angiogenesis is a crucial factor in cancer progression. Neovascularization, including tumor angiogenesis, is classified in a four-step process: 1) the basement membrane in tissues is injured locally, this induces immediate destruction and hypoxia; 2) endothelial cells are activated by angiogenic factors migrate; 3) endothelial cells proliferate and stabilize; 4) angiogenic factors continue to influence the angiogenic process (Denekamp 1993). Angiogenesis is induced when tumor tissues need nutrients and oxygen. Angiogenesis is regulated by both activator and inhibitor molecules. In fact, tumor angiogenesis is induced by up-regulation of the activity of angiogenic factors and down-regulation of inhibitors of vessel growth (Dameron et al 1994). In this perspective different molecular pathways can be targeted to stop tumor angiogenesis process by phytochemicals. For instance, phytochemicals derived from Ginseng (Panax ginseng) have been reported to inhibit tumor angiogenesis, as well as the invasion and metastasis of various types of cancer cells.

**Mitotic process**

**Invasion & Metastasis**

Phytochemicals are also promising in reducing cancer cell progression to metastasis. Metastasis is a complex process that involves cancer cell migration, invasion, dissemination through the lymphatics or vasculature, and, ultimately, colonization. Thus, phytochemicals could be used like molecularly targeted anti-metastasis 46. Phytochemicals are able to block the molecular pathways involved in metastatic events. During the Epithelial-mesenchymal transition (EMT), cancer cells acquire properties of motility and invasiveness by loss of the epithelial phenotype. Two proteins are expressed in metastasis process during EMT phase: E-cadherin that is related to epithelial-specific proteins and N-cadherin involved in the gain of mesenchymal properties by cancer cells. 47, 48 In fact, in vitro exposure of cancer cells to phytochemicals (eg EGCG,50 curcumin,51,52 and 53 resveratrol,54) showed to induce increased expression of E-cadherin and therefore decreases the mesenchymal phenotype. These phytochemicals were observed to inhibit several EMT pathways, but commonly function through inhibition of receptor and non-receptor tyrosine kinases (ERK, Src, PI3K, etc.). Remodeling of the extracellular matrix plays a crucial role in the acquisition of the phenotypic EMT by cancer cells to become invasive. This is supported by increased expression and activation of matrix metalloproteinase (MMPs) in the tumor microenvironment that are able to degrade the extracellular matrix and basement membrane. In this way MMPs activity allows cancer cell dissemination to distant sites. Different phytochemicals inhibit MMP expression and function as suppressing agents in tertiary chemoprevention. For istance, two phytochemicals: gingerol and luteolin, among many other flavonoids, appear to reduce MMP expression in pancreatic and colon cancer cells, respectively.53,56 Thus, chemoprevention may be important not only in blocking the initiation of cancer but also in reversing cancer progression.

**Quercetin**

Food-derived flavonoid quercetin, widely distributed in onions, apples, and tea, is able to inhibit growth of various cancer cells indicating that this compound can be considered as a good candidate for anticancer therapy. Quercetin is abundant in vegetables and fruits [30] usually in glycosylated, for example in form of rutin [11]. However, a study analyzed the inhibition of quercetin vs rutin on azoxymethane-induced colorectal carcinogenesis in rats, supporting that quercetin probably represents the active form inhibiting cancer respect to rutin [11]. In fact quercetin is one of the most promising phytochemicals that could be use for cancer chemoprevention.

Efficacy against several cancer types was demonstrated in both *in vitro* and *in vivo* assays and different epidemiological studies demonstrated its effect against lung cancer [31]. Quercetin has an antioxidant effect as well as modulating different intracellular signalling cascades. In fact quercetin possesses free radicals scavenging *antioxidant activity* and several anticancer effects including *anti-mutagenic, and anti-proliferative activities*, regulating several cell-signaling pathways, cell cycle, and apoptosis [11]. Moreover, quercetin is able to inhibit the Hedgehog signaling, PI3K/Akt survival signaling pathway, COX-2 and PGE2 production [32]. Quercetin is involved in the sensitization of prostate, colon, and lung cancer cells to TRAIL-induced cytotoxicity due to the upregulation of DR5 and downregulation of surviving [33].

Interestingly, the use of quercetin together with EGCG produced an enhanced effect on the inhibition of prostate cancer stem cell characteristics and epithelial-mesenchymal transition (EMT), involved in invasion and metastasis [18]. Cytotoxicity constants of quercetin measured in various human malignant cell lines of different origin were compiled from literature and a clear cancer selective action was demonstrated [ref.]. The most sensitive malignant sites for quercetin revealed to be cancers of blood, brain, lung, uterine, and salivary gland as well as melanoma whereas cytotoxic activity was higher in more aggressive cells compared to the slowly growing cells showing that the most harmful cells for the organism are probably targeted. The extrapolation in humans of the daily doses of quercetin producing an antitumor effect should be better investigated, but it certainly couldn't be achievable by dietary intake alone, thus suggesting supplementation may be necessary (ref).

**Cancer immune evasion**

In the last years different studies discovered molecular mechanisms related to the immune activation that protect against the onset of tumor cells. These discoveries have revolutionized the field of immunotherapy research. The dysfunction of the host's immune system is one of the most important mechanisms by which tumors suppress immunosurveillance. The responsible factors of immune evasion are related to T cell anergy, the presence of regulatory T cells, and systemic defects of dendritic cells derived from tumor patients. Immunosurveillance suppression involve the following factors: resistance to apoptosis, secretion of immunosuppressive cytokines, reduced expression of major histocompatibility complex (MHC) class I antigens and immunomodulatory molecules. Both host- and tumor-related mechanisms can lead to a failure to produce an effective anti-tumor-specific immune response, and these are often key factors in limiting the success of cancer immunotherapy.

*Epigallocatechin-3-gallate (EGCG)* – Among catechins, (-)-Epigallocatechin-3-gallate (EGCG), particularly abundant in green tea (*Camellia sinensis*) [13]. Indeed, the frequent consumption of green tea has been associated to a plethora of health benefits, such as the reduction in the risk of chronic diseases (e.g., cardiovascular disease, hypertension, diabetes, and cancer. These benefits have been suggested to be correlated to high content of EGCG. Various studies demonstrated that EGCG is an effective natural substance for cancer prevention and therapy, both alone and in combination with other antitumor drugs or phytochemicals. Indeed, EGCG strong antitumor activity on several cancer type has been reported (Table 3), including anaplastic thyroid carcinoma, one of the most aggressive cancer in humans (ref). On the other hand, evidences support the hypothesis that, to show chemopreventive activity, EGCG requires, in most cases, high concentrations, ranging between 0.1 μM and 0.3 μM in plasma [14]. Recently, new methods of EGCG administration have been developed to increase bioavailability and improve efficacy [ref Varoni Nutrients], they included more effective systems of delivery of EGCG by different means of administration. . ...
EGCG acts on a number of cancer-related pathways (Tab. 3), most related to the following processes: cancer cell proliferation, apoptosis, immune evasion, invasion, metastasis, and angiogenesis. The *anti-angiogenic effect* of EGCG against tumor-associated endothelial cells and endothelial progenitor cells [15] is related to inhibition of the vascular endothelial growth factor (VEGF)/VEGF receptor (VEGFR) axis [16]. In addition, EGCG is also able to *reduce cell migration*, acting on hepatocyte growth factor (HGF)/c-Met signaling [17], thus with a putative effect in decreasing rate of cancer metastasis. This *in vitro* effect was more evident when EGCG was used in addition to antitumor drugs, namely raloxifen or bortezomib, in perspective of the adjuvant therapy. Nonetheless, a synergistic effect could be observed when EGCG was tested in addition to other phytochemicals, such as quercetin [18]. In terms of *immune response modulation,* EGCG is involved in increasing immune surveillance against tumour, down-regulating the expression of indoleamine 2,3-dioxygenase, in turn involved in the anti-proliferative and pro-apoptotic effect on T cells in the cancer microenvironment [19].

*Lycopene -* Lycopene is the most promising carotenoid with effects related to cancer prevention and therapy. Moreover lycopene plays an important role in regulating hormone action, cell cycle, apoptosis, gap-junction communication and epigenetics, suggesting that their antioxidant activity may not be solely responsible for their anticancer effect [59].

An inverse correlation between lycopene intake and prostate cancer risk was observed [75]. Lycopene administered for 3 months (10 mg/day) in patients with prostate cancer decreased PSA, tumor grade and urinary tract symptoms [60]. Moreover, lycopene is able to inhibit other cancer types (Table 3). Lycopene treatment induces an alteration of serum concentrations of components of the IGF system in both prostate and colorectal cancer patients [61, 62]. In vitro, lycopene is able to modify the functionality of different pathways (Table 3). In epidemiological studies, lycopene reveals to be a promising agent in cancer therapy, especially for prostate cancer. A recent study reported that lycopene administration increased the antitumor activity of docetaxel in castration-resistant prostate tumor models [63]. Furthermore, some studies did not demonstrate an inverse correlation between lycopene consumption and cancer risk [64, 65]. This aspect could be explained, in part, by interindividual variations with regard to genetic background. In fact, polymorphisms in the genes XRCC1 and MNSOD were reported to determine the effect of lycopene regarding prostate cancer risk [66, 67]. Finally, an epidemiological study in elderly Americans indicated that high tomato intake was associated with a 50% reduction in mortality from cancers at all sites, and a case-control study in Italy showed potential protection of high consumption of lycopene from tomatoes against cancers of the digestive tract [75].

**Targeting of cancer stem cells**

In the last years different studies revealed that cancer is an epigenetic dysfunction [87]. In fact, phytochemicals are able to affect normal cell growth, proliferation and differentiation and also to revert cancer related epigenetic dysfunctions, reducing tumorigenesis, preventing metastasis and/or increasing chemo and radiotherapy efficacy [88]. A crucial role for tumor resistance to chemio and radiotherapy and tumor relapse is played by cancer stem cells (CSCs) that reside in specific hypoxic and acidic microenvironments or niches [89, 90]. In this perspective cancer stem cells could be a target of phytochemicals to modulate the hypoxic and acidic tumor microenvironment that has an important role in cancer progression.

Different studies have been found that CSCs have various self-renewal related signaling pathways that are deregulated in many cancer types [91, 92]. In this perspective to limit the role of cancer stem cells a potential strategy could be by inducing CSCs into differentiation and thus reducing their stemness related phenotype [93, 94]. Plant-derived compounds have an important role in the regulation/inhibition of CSC self-renewal. Various CSC self-renewal related pathways are regulated by phytochemicals such as Wnt/β-catenin, Hedgehog and Notch [95]. Phytochemicals that have been described to either directly or indirectly affect these self-renewal signaling pathways are as follows: curcumin (CUR), genistein (GEN), sulforaphane (SFN), indole-3-carbinol and 3,3′-diindolylmethane, epigallocatechin-3-gallate (EGCG), resveratrol (RES), lycopene and piperin. In fact the previous phytochemicals listed contribute to the physiological regulation of normal (non tumorigenic) stem cells and also to the reduction of CSC growth [95].

## Conclusion

According to current literature, the most effective plant-based anti-cancer agents are: camptothecin (from *Camptotheca acuminate Decne*), vinblastine and vincristine (from *Catharanthus roseus*), or paclitaxel (from *Taxus brevifolia*) [3].

In this article we reviewed the anticancer activity of a group of phytochemicals representing good candidates for chemopreventive and chemotherapeutic applications. Phytochemicals have a relative pharmacological safety, as confirmed by different studies [ref.]. Moreover, they exert their anticancer effects through multiple target molecular pathways. Different epidemiological studies enhanced that high dietary consumption of vegetables and fruits reduced the risk of cancer onset.

The stages of cancer development are regulated by several signaling pathways, and phytochemicals can act on these molecular pathways with the aim of multitargeted therapies. This article has reviewed the most important molecular targets of phytochemicals. Phytochemicals could be used like supplements for the management of cancer prevention and treatment. The additive and synergic activity of phytochemicals could be combined with chemotherapy and radiotherapy, limiting the doses and trimming down the toxicity.

**Future perspectives**

It is now possible to discover the specific molecular mechanisms responsible for the antitumor effects of traditionally used phytochemicals, thanks to the advanced technologies currently available in scientific research.

Specific preclinical investigations will allow researchers to determine the scientific basis upon which more targeted human studies can be designed (Ref Varoni Pharm Biol). A phytochemical should be characterized regarding its effect on gene and protein expression, also affected by the genetic background of each individual. The potential effects of different polymorphisms on the activity of selected compound can be considered. The chemopreventive and therapeutic effects could vary according to specific patients (refs). In the context of a "personalized medicine", this approach will allow outcomes that are more predictable. Moreover, the identification of biomarkers to differentiate between responders and non-responders may allow to early comprehension of pharmacological effects of phytochemicals (i.e., dose and temporal changes in cellular small-molecular-weight compounds) [73].

Nonetheless, variations in the molecular targets (receptors and/or signal transducers) of the phytochemical play also a pivotal role. More detailed studies need to be performed in order to better understand how each phytochemical can be of benefit in the prevention and therapy of specific tumor types.

In addition, pharmacokinetic and pharmacodynamic of specific compounds should be better investigated. In fact bioavailability of the phytochemicals plays a key-role for their effectiveness. The results of such analyses are essential to determine the pharmacokinetic/pharmacodynamic profiles of the compound, as well as to confirm the interactions with other molecules in the context of adjuvant therapy. Thus, an appropriate combination therapy will potentially lead to a reduction in side effects without modifying or even increasing the chemotherapeutic effect. So phytochemicals are promising to develop new strategies for cancer chemotherapy.

Moreover, the safety and low cost of these compounds make them promising molecules for cancer prevention, particularly in subjects at increased risk of cancer development due to their genetic background or unavoidable and long-term exposure to carcinogens.